\documentclass[aps,prb,showpacs,twocolumn,superscriptaddress,floatfix]{revtex4}

\usepackage{graphicx}
\usepackage{dcolumn}
\usepackage{bm}
\usepackage{amsmath}
\usepackage{amssymb}

\begin{document}

\renewcommand{\vec}[1]{\mbox{\boldmath $#1$}}


\title{Molecular spin-orbit excitations in the $J_{\rm eff}=1/2$ frustrated spinel GeCo$_2$O$_4$}


\author{K. Tomiyasu}
\email[Electronic address: ]{tomiyasu@m.tohoku.ac.jp}
\affiliation{Department of Physics, Tohoku University, Aoba, Sendai 980-8578, Japan}
\author{M. K. Crawford}
\affiliation{Central Research and Development Department, DuPont, 
Wilmington, Delaware 19880, USA}
\author{D. T. Adroja}
\affiliation{ISIS Facility, Rutherford Appleton Laboratory, Chilton, Didcot, OX11 0QX, UK}
\author{P. Manuel}
\affiliation{ISIS Facility, Rutherford Appleton Laboratory, Chilton, Didcot, OX11 0QX, UK}
\author{A. Tominaga}
\affiliation{Department of Physics, Chuo University, Bunkyo, Tokyo 101-8324, Japan}
\author{S. Hara}
\affiliation{Department of Physics, Chuo University, Bunkyo, Tokyo 101-8324, Japan}
\author{H. Sato}
\affiliation{Department of Physics, Chuo University, Bunkyo, Tokyo 101-8324, Japan}
\author{T. Watanabe}
\affiliation{Department of Physics, CST, Nihon University, 
Chiyoda, Tokyo 101-8308, Japan}
\author{S. I. Ikeda}
\affiliation{Nanoelectronics Research Institute, National Institute of AIST, 
Tsukuba 305-8568, Japan}
\author{J. W. Lynn}
\affiliation{NIST Center for Neutron Research, Gaithersburg, Maryland 20899, USA}
\author{K. Iwasa}
\affiliation{Department of Physics, Tohoku University, Aoba, Sendai 980-8578, Japan}
\author{K. Yamada}
\affiliation{WPI AIMR, Tohoku University, 
Aoba, Sendai 980-8577, Japan}


\date{\today}

\begin{abstract}
We describe powder and single-crystal inelastic neutron scattering experiments on a spinel-type antiferromagnet GeCo$_2$O$_4$, represented by an effective total angular momentum $J_{\rm eff}=1/2$. Several types of non-dispersive short-range magnetic excitations were discovered. The scattering intensity maps in $\vec{Q}$ space are well reproduced by dynamical structure factor analyses using molecular model Hamiltonians. The results of analyses strongly suggest that the molecular excitations below $T_N$ arise from a hidden molecular-singlet ground state, in which ferromagnetic subunits are antiferromagnetically coupled. The quasielastic excitations above $T_N$ are interpreted as its precursor. A combination of frustration and $J_{\rm eff}=1/2$ might induce these quantum phenomena.
\end{abstract}


\maketitle

%
\section{\label{sec:Introduction}INTRODUCTION}

Since their initial proposals,~\cite{Pauling_1935, Wannier_1950, Anderson_1956} the concept of geometrical spin frustration has been intensively studied. Geometrical frustration has been shown to give rise to novel forms of spin-liquid-like fluctuations in a paramagnetic phase, such as spin molecules, spin ices, and spin vortices.~\cite{Ballou_1996,Lee_2002,Yasui_2002,Nakatsuji_2005,Kawamura_2010} Recently, dynamical spin molecules were discovered as non-dispersive excitation modes within a magnetically ordered phase, where frustration was assumed to be relieved by a lattice deformation.~\cite{Tomiyasu_2008} 

Meanwhile, it was demonstrated that an effective total angular momentum $J_{\rm eff}=1/2$, generated by a spin-orbit coupling (SOC), provides a new playground for correlated electrons. For example, a Mott instability with spin-orbit integrated narrow band was confirmed in Sr$_2$IrO$_4$, and a quantum spin-hall effect at room temperature was theoretically predicted for Na$_2$IrO$_3$.~\cite{Kim_2008,Kim_2009,Shitade_2009} These iridates possess Ir$^{4+}$ with low-spin $(t_{2g})$$^5$ configuration, of which the ground states are described by $J_{\rm eff}=1/2$ with unquenched orbital angular momentum ($L=1$). The $L=1$ states are related to $t_{2g}$ triplets ($xy,yz,zx$): $|L^{z}=\pm1\rangle = (|yz\rangle \pm i |zx\rangle)/\sqrt{2}$ and $|L^{z}=0\rangle = |xy\rangle$. The value of 1/2 and the complex orbitals of $J_{\rm eff}$ are expected to enhance the quantum nature accompanied with orbital degree of freedom.~\cite{Shitade_2009}

Then, an interest in the combination of frustration and $J_{\rm eff}=1/2$ will naturally arise. In fact, there are reports of Ir$^{4+}$ systems, {\it e.g.} a face-centered cubic system K$_2$IrCl$_6$ forming a magnetic complex IrCl$_6$ with remarkably mixed orbital, a hyperkagome system Na$_4$Ir$_3$O$_8$ with quantum spin liquid, and pyrochlore systems $Ln_2$Ir$_2$O$_7$ ($Ln$=Nd, Sm, Eu) with metal-insulator transition.~\cite{Lynn_1976,Okamoto_2007,Matsuhira_2007,Sakata_2011} However, an extremely strong neutron absorption of Ir nuclei ($\sim425$ barns for thermal neutrons)~\cite{ScattL_1992} and lack of large single crystals hamper the successful inelastic neutron scattering experiments, a prime tool for the study of magnetic frustration. 

\begin{figure}[htbp]
\begin{center}
\includegraphics[width=0.99\linewidth, keepaspectratio]{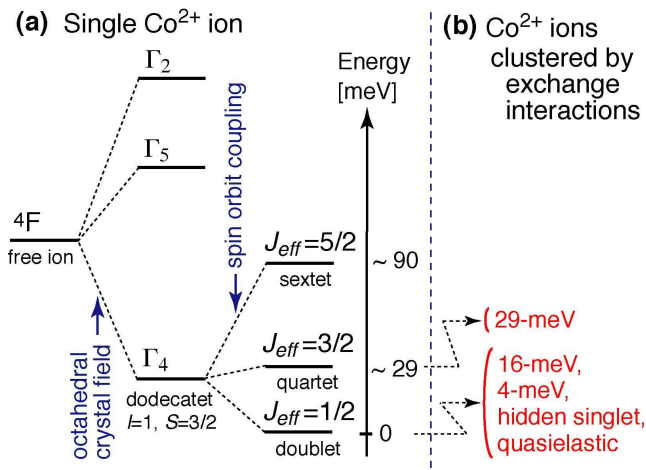}
\end{center}
\caption{\label{fig:schemes} (Color online) (a) Energy level scheme of Co$^{2+}$ ion under octahedral crystal field and SOC. (b) Correspondence with the data measured in the present experiments. }
\end{figure}

The spinel-type antiferromagnet GeCo$_2$O$_4$ is a promising candidate with frustration and $J_{\rm eff}=1/2$. In this material, well-known SOC-active Co$^{2+}$ ions octahedrally surrounded by anions form a lattice of corner-sharing tetrahedra, which is geometrically frustrated, and Ge$^{4+}$ ions are nonmagnetic. 
Figure~\ref{fig:schemes}(a) shows the energy-level schemes of a single-ion state of Co$^{2+}$ ($d^7$). The crystal field and SOC yield $J_{\rm eff}=1/2$, 3/2, and 5/2 states with $L=1$ and $S=3/2$.~\cite{Kanamori_1957,Lines_1963,Lashley_2008,Watanabe_2008} Antiferromagnetic order with propagation vector $\vec{q}_{m}=(1/2,1/2,1/2)$ and a tiny tetragonal lattice deformation ($c/a\simeq1.001$) simultaneously occur at $T_{N}\simeq21$ K, which is suppressed compared to the Curie-Weiss temperature $\theta_{W}\simeq81$ K.~\cite{Hubsch_1987,Diaz_2004,Diaz_2006,Hoshi_2007} Spin-liquid-like fluctuations above $T_N$ (quasielastic mode) and a non-dispersive magnetic excitation mode below $T_N$ (4-meV mode) were also found by powder inelastic neutron scattering.~\cite{Lashley_2008} 

%

In this paper, we comprehensively study magnetic excitations above and below $T_N$ in GeCo$_2$O$_4$ in wide momentum ($\vec{Q}$) and energy ($E$) ranges by powder and single-crystal inelastic neutron scattering. The experimental results and numerical analyses strongly suggest manifestation of highly-frustrated quantum states in this cobaltite. 

%
\section{\label{sec:Experiments}EXPERIMENTS}

Initial single crystal studies were performed at the NIST Center for Neutron Research using the BT-2 and BT-9 triple axis spectrometers. Single-crystal inelastic neutron scattering experiments were performed on the triple axis spectrometer TOPAN, installed at the JRR-3 reactor, JAEA, Tokai, Japan. The final energy of the neutrons was fixed at $E_{f}=13.5$ meV with horizontal collimation sequence of blank-100$^{\prime}$-100$^{\prime}$-blank. A sapphire filter and a pyrolytic graphite filter efficiently eliminated fast neutrons and the higher order contamination, respectively. 
Single-crystal rods of GeCo$_2$O$_4$ were grown by a floating zone method. Details of the crystal growth are summarized in Ref~\cite{Hara_2005}. The rod size was about 4 mm diameter and 30 mm height. The three co-aligned single crystals were enclosed with He exchange gas in an aluminum container, which was placed under the cold head of a closed-cycle He refrigerator. 

Powder inelastic neutron scattering experiments were performed on the direct geometry chopper spectrometer HET, installed at the spallation neutron source, ISIS Facility, UK. The energy of the incident neutrons was fixed at $E_{i}=59$ and 29 meV. A 35 g powder specimen of GeCo$_2$O$_4$ was synthesized by a solid state reaction method, filled in an envelope made from thin aluminum foil, and inserted in a refrigerator with He exchange gas. 

%
\section{\label{sec:Results}RESULTS}

%
\begin{figure*}[htbp]
\begin{center}
\includegraphics[width=0.95\linewidth, keepaspectratio]{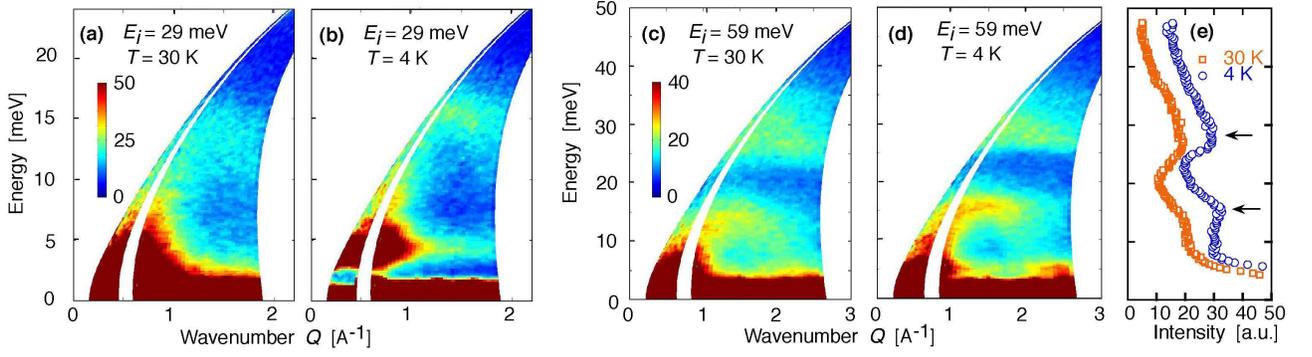}
\end{center}
\caption{\label{fig:pwd} (Color online) (a)-(d) Color images of powder inelastic neutron scattering data with different incident energies and temperatures. The color tones indicate the scattering intensity in mbarn/(sr$\cdot$formula) units. (e) Energy spectra, averaged from 3 to 29 deg in scattering angle ($Q=0.3$ to 2.7 {\AA}$^{-1}$ for elastic condition) in (c) and (d). The arrows indicate the 16-meV and 29-meV modes. }
\end{figure*}
\begin{figure*}[htbp]
\begin{center}
\includegraphics[width=0.9\linewidth, keepaspectratio]{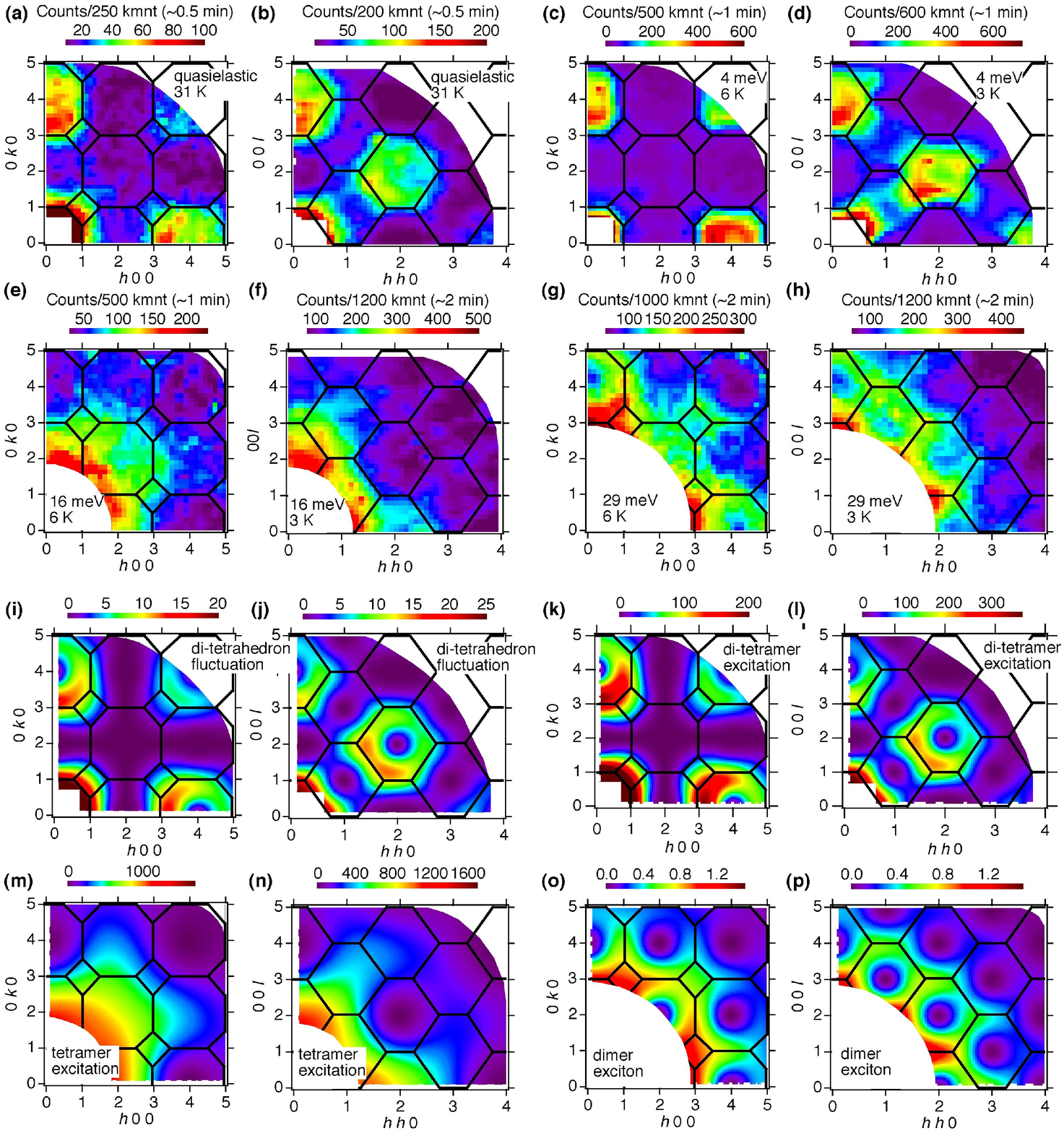}
\end{center}
\caption{\label{fig:sx} (Color online) (a)-(h) Color images of single-crystal inelastic neutron scattering data, measured in the $hk$0 and $hhl$ zones in a constant energy scan mode. (a) and (b) were measured at $E=4$ meV. (i)-(p) One-to-one correspondence between calculated patterns as identified by the molecular models shown in Figs.~\ref{fig:models}(a) to \ref{fig:models}(e) and described in the text. The bold lines show the Brillouin zone boundary of the spinel structure. For the calculated patterns, the horizontal bars indicate the scattering intensity in arbitrary units. }
\end{figure*}
%

Figures~\ref{fig:pwd}(a) and \ref{fig:pwd}(b) show the powder data with $E_{i}=29$ meV. Above $T_N$ the quasielastic mode is observed around $Q \equiv |\vec{Q}| = |\vec{q}_{m}| = 0.66$ {\AA}$^{-1}$, as shown in Fig.~\ref{fig:pwd}(a). Below $T_N$ spin-wave-like dispersion rises up from around $Q=|\vec{q}_{m}|$ in addition to the previous discovered 4-meV mode,~\cite{Lashley_2008} as shown in Fig.~\ref{fig:pwd}(b). 
Figures~\ref{fig:pwd}(c) and \ref{fig:pwd}(d) show the data with $E_{i}=59$ meV. Two discrete levels are discovered around $E=16$ and 29 meV both above and below $T_N$, indicating that the two modes are not spin waves. Below $T_N$ these modes slightly sharpen and harden. 

We measured $\vec{Q}$ correlations of the quasielastic mode above $T_N$ and the 4-meV, 16-meV, and 29-meV modes below $T_N$ in a constant-$E$ scan mode by single-crystal inelastic neutron scattering, as shown in Figs.~\ref{fig:sx}(a) to \ref{fig:sx}(h). The scattering intensity distributions with characteristic patterns decrease at higher $Q$, as expected for the Co magnetic form factor, indicating that the excitations must be attributed to magnetic origin and not phononic. 
Figures~\ref{fig:sx}(a) and \ref{fig:sx}(b) show the data for the quasielastic mode, measured at $E=4$ meV. The intensity is strong only in the 400, 440, and 222 Brillouin zones, and is distributed near the edges of the zones. 
Figures~\ref{fig:sx}(c) and \ref{fig:sx}(d) show the data measured at 4 meV below $T_{N}$. Though it is difficult to remove the spin-wave component spread around $h/2$ $k/2$ $l/2$ reciprocal lattice points (magnetic Bragg reflection points for elastic scattering), the scattering pattern is quite similar to that for the quasielastic scattering. 
Figures~\ref{fig:sx}(e) to \ref{fig:sx}(h) show the data for the 16-meV and 29-meV modes. The scattering intensity of the former mode is relatively strong except for the above Brillouin zones, whereas that of the latter mode is distributed on every zone boundary. 

%
\section{\label{sec:Model_Analyses}MODEL ANALYSES}

%
\begin{figure*}[htbp]
\begin{center}
\includegraphics[width=0.9\linewidth, keepaspectratio]{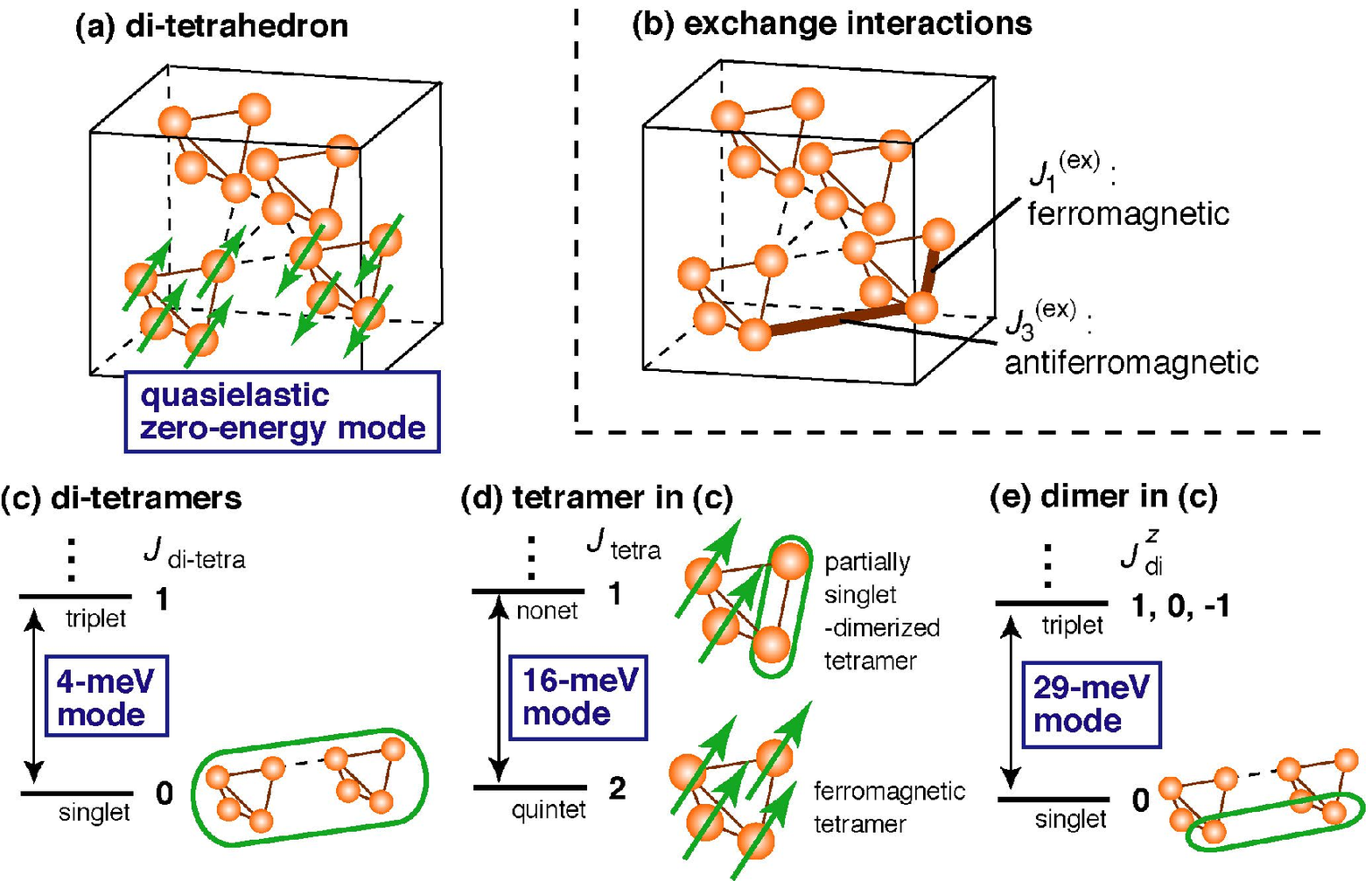}
\end{center}
\caption{\label{fig:models} (Color online) (a)(c)(d)(e) Schematic representations of the molecular models. The green arrows represent magnetic moments of the Co$^{2+}$ ion, and green ellipsoids represent a non-magnetic singlet formation. All the moments dynamically fluctuate in arbitrary directions with the relative correlations. The structural units shown in (a) and (c) are identical to each other. In (c) to (e), the representative states are depicted. (b) First and third neighbor exchange interactions. Representative bonds are shown. }
\end{figure*}
%

We analyzed the quasielastic mode using a molecular model, as for spin frustrated systems.~\cite{Lee_2002,Tomiyasu_2008} For elastic and quasielastic magnetic neutron scattering, the cross section is described by 
\begin{equation}
S(\vec{Q}) = C_{0} |F(\vec{Q})|^2 \bigl| \sum_{j=1}^{N} J_{j\perp} \exp({i \vec{Q} \cdot \vec{r}_{j}}) \bigr|^2, 
\label{eq:xs_el}
\end{equation}
where $C_{0}$ is a proportional constant of intensity, 
$F(\vec{Q})$ is the magnetic form factor of the Co$^{2+}$ ion, for which the Watson-Freeman one was used below,~\cite{Watson_1961} 
$j$ labels the site of the Co$^{2+}$, 
$N$ is total number of the sites in a molecule, 
$\vec{r}_{j}$ and $\vec{r}_{j^{\prime}}$ are those positions, 
and $J_{j\perp}$ is an expected value of $\vec{Q}$-perpendicular component in $\vec{J}_{j}$.~\cite{Marshall_1971} 
When colinear $\vec{J}$'s fluctuate in arbitrary directions like in a hexagonal-type quasielastic mode observed in the typical spin-frustrated system ZnCr$_2$O$_4$, $J_{j\perp}$ ($=S_{j\perp}$) takes on only $\pm1$.~\cite{Lee_2002} Following this treatment, we searched for and found a di-tetrahedral model for the quasielastic mode in GeCo$_2$O$_4$, as shown in Fig.~\ref{fig:models}(a). Figures~\ref{fig:sx}(i) and \ref{fig:sx}(j) show the calculated patterns, which are in good agreement with the experimental patterns of Figs.~\ref{fig:sx}(a) and \ref{fig:sx}(b). 

For inelastic magnetic neutron scattering, the cross section is described by 
\begin{multline}
S(\vec{Q},E) =  \\
C_{0} |F(\vec{Q})|^2 \delta(\hbar\omega-E)
\Biggl( \sum_{\alpha,\beta=1}^{3} (\delta_{\alpha\beta} - \frac{Q_{\alpha}Q_{\beta}}{|\vec{Q}|^2})  \phantom{0} \times \\
\sum_{j,j^{\prime}=1}^{N} 
\langle\lambda\mid \hat{J}_{j^{\prime}}^{\alpha} \mid\lambda^{\prime}\rangle
\langle\lambda^{\prime}\mid \hat{J}_{j^{\prime}}^{\beta} \mid\lambda\rangle
\exp\{i\vec{Q}\cdot(\vec{r}_{j}-\vec{r}_{j^{\prime}})\} \Biggr), 
\label{eq:xs_general}
\end{multline}
where $\alpha$ and $\beta$ are $(x,y,z)$, 
$N$ is number of sites in a molecule, 
$\mid\lambda\rangle$ and $\mid\lambda^{\prime}\rangle$ are molecular ground and excited states, respectively, 
$\hat{J}$ is a total angular momentum operator, 
and the parentheses indicates an orientational average over equivalent molecules.~\cite{Marshall_1971} One cannot generally apply Eq.~(\ref{eq:xs_el}), which is obtained from Eq.~(\ref{eq:xs_general}) only when the matrix elements $\langle\lambda^{\prime}\mid\hat{J}_{j}^{\alpha}\mid\lambda\rangle$ can be simply reduced for elastic scattering. 
In the following we try to reproduce our inelastic scattering data using a relatively simple molecular model. We assume effective molecular Hamiltonians, and numerically evaluate $\langle\lambda^{\prime}\mid\hat{J}_{j}^{\alpha}\mid\lambda\rangle$ and the cross section. The assumption of a molecular formation implies a remarkably-mixed molecular orbital, which will enhance intra-molecular exchange interactions and suppress atomic orbital characters like anisotropy.~\cite{comm_aniso} Therefore, we ignore the exchange field outside the molecule (Lorentz-like local magnetic field) and the directional term $(\delta_{\alpha\beta} - Q_{\alpha}Q_{\beta}/|\vec{Q}|^2)$ in Eq.~(\ref{eq:xs_general}). An orientational average over dynamically fluctuating molecules will also substantially suppress the directional dependence. For simplicity the atomic Watson-Freeman form factor is used for $F(\vec{Q})$ in Eq.~(\ref{eq:xs_general}) again.~\cite{Watson_1961} 

Firstly, we exactly diagonalized a tetramer Hamiltonian: 
\begin{equation}
\hat{H}_{\rm tetra}=J^{\rm (ex)}_{1}\sum_{i,j=1}^{4} \hat{\vec{J}_{i}} \cdot \hat{\vec{J}_{j}}, 
\label{eq:H_tetra}
\end{equation}
where $J_{i}=1/2$, $i$ and $j$ are positions of the tetrahedral sites (Fig.~\ref{fig:models}(d)), $\sum_{i,j}$ means summation over all $\vec{J}$ pairs (not doubly counted), and $J^{\rm (ex)}_{1}$ is a first-neighbor exchange interaction that is ferromagnetic as expected from the quasielastic mode (Fig.~\ref{fig:models}(a)) and the Goodenough-Kanamori rule.~\cite{Goodenough_1960} The 16 ($=2^4$) basis states of $|J_{1}^{z},J_{2}^{z},J_{3}^{z},J_{4}^{z}\rangle$ were used, where $J_{i}^{z}=\pm1/2$. 
Figure~\ref{fig:models}(d) shows the obtained level scheme with $J^{\rm (ex)}_{1}=-8$ meV. The ground states are described as ferromagnetic quintets with $J_{\rm tetra}=2$, and the first excited states are nonets with $J_{\rm tetra}=1$ and $E=16$ meV, where $\vec{J}_{\rm tetra}=\sum_{i=1}^{4}\vec{J}_{i}$. The nonet can generate all states with $J_{\rm tetra}^{z}=\pm1,0$ with a $J_{\rm eff}=1/2$ dimer-singlet bond by their linear combinations (e.g. Fig.~\ref{fig:models}(d)). The calculated patterns for excitation processes from the ground states to the excited states are shown in Figs.~\ref{fig:sx}(m) and \ref{fig:sx}(n), which are in excellent agreement with the experimental patterns of Figs.~\ref{fig:sx}(e) and \ref{fig:sx}(f) (16-meV). 

Secondly, we diagonalized a di-tetramer Hamiltonian:  
\begin{equation}
\hat{H}_{\rm di\mathchar`-tetra}=J^{\rm (ex)}_{ij}\sum_{i,j=1}^{8} \hat{\vec{J}_{i}} \cdot \hat{\vec{J}_{j}}, 
\label{eq:H_di-tetra}
\end{equation}
where $J^{\rm (ex)}_{ij}=J^{\rm (ex)}_{1}$ and $J^{\rm (ex)}_{3}$, $i$ and $j$ are positions of sites in the di-tetramer (Fig.~\ref{fig:models}(c)), and the 256 ($=2^8$) basis states of $|J_{1}^{z},J_{2}^{z},J_{3}^{z}, ... ,J_{8}^{z}\rangle$ were used. The sign of $J^{\rm (ex)}_{3}$ is antiferromagnetic, being consistent with the quasielastic mode and previous neutron diffraction reports.~\cite{Diaz_2006}
Figure~\ref{fig:models}(c) shows the level scheme with $J^{\rm (ex)}_{1}=-8$ meV and $J^{\rm (ex)}_{3}=10$ meV. The ground state is described as a non-magnetic singlet with $J_{\rm di\mathchar`-tetra}=0$, and the first excited states are triplet with $J_{\rm di\mathchar`-tetra}=1$, where $\vec{J}_{\rm di\mathchar`-tetra}=\sum_{i=1}^{8}\vec{J}_{i}$. Figures~\ref{fig:sx}(k) and \ref{fig:sx}(l) show the calculated patterns of the singlet-triplet excitations, which are similar to those for the quasielastic mode (Figs.~\ref{fig:sx}(i) and \ref{fig:sx}(j)), and are identified as the 4-meV mode. 

For the 29-meV mode, we could find no model within $J_{i}=1/2$ after many trials. On the other hand, interestingly, other cobalt compounds KCoF$_3$, CoO, and La(Sr)$_2$CoO$_4$, consisting of Co$^{2+}$ ions octahedrally surrounded by anions as well, exhibit excitations around 30 meV.~\cite{Holden_1971,Buyers_1971,Tomiyasu_2006,Helme_2009} These excitations are interpreted as the lowest-energy SOC excitations ({\it i.e.} excitons).~\cite{Holden_1971,Buyers_1971,Tomiyasu_2006,Helme_2009} In analogy with these cobalt compounds, the 29-meV mode in GeCo$_2$O$_4$ is to be excitons. 

Thus, thirdly, we studied molecular excitons from $J_{\rm eff}=1/2$ to $J_{\rm eff}=3/2$,  assuming the following Hamiltonian of third-neighbor dimer (Fig.~\ref{fig:models}(e)):  
\begin{equation}
\hat{H}_{\rm di}=J^{\rm (ex)\prime}_{3}\sum_{i,j=1}^{2} \hat{\vec{S}}_{i} \cdot \hat{\vec{S}}_{j}, 
\label{eq:H_tri}
\end{equation}
where $J^{\rm (ex)\prime}_{3}=2$ meV is an effective value of $J^{\rm (ex)}_{3}=10$ meV for spin-3/2 estimated by a relation $J^{\rm (ex)\prime}_{3}S(S+1)=J^{\rm (ex)}_{3}J_{\rm eff}(J_{\rm eff}+1)$, and $i$ and $j$ label sites in this dimer. This Hamiltonian is expressed by $\vec{S}$, not $\vec{J}$, for the cobalt compounds.~\cite{Holden_1971,Buyers_1971,Tomiyasu_2006,Helme_2009} The 36 ($=6^2$) basis states of $|J_{1},J_{1}^{z}\rangle \otimes |J_{2},J_{2}^{z}\rangle$ were used, where $|1/2,\pm1/2\rangle$ single-ion states have zero energy, and $|3/2,\pm3/2\rangle$ and $|3/2,\pm1/2\rangle$ ones have a 29 meV SOC excitation energy for $|J_{i},J_{i}^{z}\rangle$. Figures~\ref{fig:sx}(o) and \ref{fig:sx}(p) show the calculated patterns, which take into account the processes from the $J_{\rm eff}=1/2$ ground dimer-singlet to the first excited triplet with $J_{\rm eff}=3/2$. The calculated patterns are in good agreement with the experimental patterns (Figs.~\ref{fig:sx}(g) and \ref{fig:sx}(h)). 

More precisely, the Co$^{2+}$ feels an additional trigonal component of crystal electric field, which keeps the $J_{\rm eff}=1/2$ ground doublet but splits the $J_{\rm eff}=3/2$ quartet into two Kramers doublets ($J_{\rm eff}^{z}=\pm3/2$ doublet and $J_{\rm eff}^{z}=\pm1/2$ one) in the level scheme shown in Fig.~\ref{fig:schemes}(a).~\cite{Lashley_2008,Lines_1963} In fact, as shown in Fig.~\ref{fig:pwd}(e), the experimental spectrum around 29 meV is asymmetrically spread up to 40 meV, suggesting this splitting. However, the profile is too broad to clearly resolve into the two levels. Therefore we carried out the above calculation integrating the $J_{\rm eff}=3/2$ quartet. We also confirmed that the transitions to each Kramers doublets give the same patterns.

In this way, we identified the quasielastic mode as an antiferromagnetic di-tetrahedral cluster (Fig.~\ref{fig:models}(a)), consisting of Co$^{2+}$ ions with $J_{\rm eff}=1/2$. Furthermore, assuming model Hamiltonians, we assigned the 4-meV to the singlet-triplet excitations in a di-tetramer with the same structural unit (Fig.~\ref{fig:models}(c)), the 16-meV to quintet-nonet excitations in the one ferromagnetic tetramer (Fig.~\ref{fig:models}(d)), and the 29-meV to SOC excitons from $J_{\rm eff}=1/2$ to 3/2 in a third-neighbor-distant antiferromagnetic dimer (Fig.~\ref{fig:models}(e)). All the excitations can be regarded as intra-activations of the di-tetramer. The correspondence relation between excitations and $J_{\rm eff}$ states are shown in Fig.~\ref{fig:schemes}(b). 

%
\section{\label{sec:Discussion}DISCUSSION}

We discuss the ferromagnetic tetramer (Fig.~\ref{fig:models}(d)). The remarkable spatial confinement of magnetic correlation demonstrates the existence of frustration. But frustration is normally based on antiferromagnetism. So what is frustrated in GeCo$_2$O$_4$? One factor will be the frustration among $J^{\rm (ex)}_{1}$, $J^{\rm (ex)}_{3}$, and the other exchange interactions. Aside from this, however, an orbital system is inherently frustrated even on a simple cubic lattice; when orbitals (directions of the electron cloud) are arranged to gain bond energy for one direction, this configuration is not fully favorable for other bonds.~\cite{Khomskii_2003,Tanaka_2007,Nasu_2008} GeCo$_2$O$_4$ also has an orbital angular momentum, which is a kind of orbital, and is in the geometrically frustrated pyrochlore lattice. Therefore, both exchange and orbital frustration likely coexist in GeCo$_2$O$_4$. 

Next we discuss the di-tetramer. According to the above analyses, the di-tetramer singlet ground state is surprisingly hidden as origin of the molecular excitations below $T_N$.  Indeed, a singlet formation is an effective way to suppress frustration and degree of freedom. However, the formation does not necessarily mean that all the magnetic moment disappears, because the $g$ factor is arbitrary in our analyses, being consistent with the coexistence of singlet and magnetic order. This partial-singlet model can explain why GeCo$_2$O$_4$ exhibits magnetic order with only about 3 $\mu_{\rm B}$ per Co$^{2+}$,~\cite{GCO_unpb} which is 1 $\mu_{\rm B}$ lower than a normal value of 4 $\mu_{\rm B}$ generated by SOC like in CoO.~\cite{Kanamori_1957} 

It should be noted that a typical spin-frustrated spinel antiferromagnet MgCr$_2$O$_4$ (Cr$^{3+}$, $d^3$, $S=3/2$) similarly exhibits a set of quasielastic modes above $T_N$ (hexamer) and a gapped non-dispersive excitation mode below $T_N$, of which the scattering intensity distributions in $\vec{Q}$ space are the same.~\cite{Tomiyasu_2008} In addition, MgCr$_2$O$_4$ exhibits magnetic order with only 2.2 $\mu_{\rm B}$,~\cite{Shaked_1970} which is about 1 $\mu_{\rm B}$ lower than the full moment 3 $\mu_{\rm B}$. Therefore, a hexamer-type singlet ground state would give rise to both the gapped excitation mode and the partial disappearance of the magnetic moment below $T_N$. We also remark that the 1 $\mu_{\rm B}$ decrease is observed in the isomorphic systems ZnCr$_2$O$_4$ and HgCr$_2$O$_4$.~\cite{Ji_2009,Matsuda_2007}

The di-tetramer can be energetically regarded as a dimer of the rigid tetramers with $J_{\rm tetra}=2$; binding energy in a ferromagnetic tetramer ($\sim36$ meV = $6J^{\rm (ex)}_{1}\cdot J_{\rm eff}(J_{\rm eff}+1)$) is higher than antiferromagnetic coupling energy between the two tetramers ($\sim24$ meV = $(4J^{\rm (ex)}_{3}+J^{\rm (ex)}_{1})\cdot J_{\rm eff}(J_{\rm eff}+1)$). We also numerically confirmed that the $J_{\rm tetra}$-2 dimer has a ground singlet with the combination of $J_{{\rm tetra},i}^{z}=\pm2,\pm1,0$ and the first excited triplets within the 25 ($=5^2$) basis states of $|J_{\rm tetra,1}^{z},J_{\rm tetra,2}^{z}\rangle$. This extended-dimer picture naturally gives us the interpretations of the 4-meV mode as a localized singlet-triplet excitation and of the quasielastic mode as its precursor fluctuations, as observed in the frustrated spin-1/2 system SrCu$_2$(BO$_3$)$_2$  with the two-dimensional Shastry-Sutherland lattice.~\cite{Kageyama_2000} 

Since its introduction as a mechanism for high-temperature superconductivity, dimer-based quantum cooperative phenomena like resonating valence bond (RVB) and valence bond solid (VBS) have been sought after in fields of magnetism and strongly correlated electron systems.~\cite{Anderson_1987} SrCu$_2$(BO$_3$)$_2$ is one of the great successes. In contrast to the borate, the molecular formations in GeCo$_2$O$_4$ are characterized by the existence of a ferromagnetic molecule, $J_{\rm eff}=1/2$, and the three-dimensional pyrochlore lattice with almost regular triangles. In this sense, GeCo$_2$O$_4$ could be positioned as a new class of quantum cooperative systems caused by frustration and $J_{\rm eff}=1/2$. 

We list two other intriguing characters of frustration and $J_{\rm eff}=1/2$. One character will be the fact that all the molecular excitations involve aspects of not only spin but also orbital excitations by SOC (molecular orbitons). Another character will be the emergence of molecular excitons (29-meV). Excitons normally appear within a single atom or ion with SOC, and are occasionally propagated with very narrow dispersion width ($\sim0.5$ meV) like in a 4$f$ electron system.~\cite{Kuwahara_2005} Furthermore, the 3$d$ electron cobalt systems exhibit more dispersive excitons around 30 meV (over 5 meV width), propagated by stronger exchange interactions than in 4$f$ systems.~\cite{Holden_1971,Buyers_1971,Tomiyasu_2006,Helme_2009} However, these excitons are molecular, which are locally collective but are not propagated. 

%
\section{\label{sec:Conclusions}CONCLUSIONS}

We discovered several types of non-dispersive short-range excitations in a three-dimensional frustrated GeCo$_2$O$_4$ with $J_{\rm eff}=1/2$ by powder and single-crystal inelastic neutron scattering. The scattering intensity maps in $\vec{Q}$ space are well reproduced by quantum-mechanical molecular models. The model analyses strongly suggest that a molecular-singlet ground state consisting of ferromagnetic sub-molecules is hidden below $T_N$, which gives origin to the molecular excitations. The quasielastic excitations above $T_N$ are interpreted as a precursor of this quantum ground state. The spin and orbital frustrations of $J_{\rm eff}$ lead to the molecular-singlet formation and the ferromagnetic molecule one, respectively. Further experimental and theoretical works will be needed to fully elucidate this hidden molecular partial-singlet conjecture and clarify the molecular orbital formations. 

\acknowledgments
We thank Mr. K. Nemoto and Mr. T. Asami for their supports in JAEA, Mr. M. Onodera for his supports in Tohoku university, and Professors Y. Motome and S. Ishihara for fruitful discussion. The neutron experiments in JAEA were performed under User Programs conducted by ISSP, University of Tokyo. This work was supported by the MEXT of Japan, Grants in Aid Young Scientists (B) (22740209), Priority Areas (22014001), Scientific Researches (S) (21224008) and (A) (22244039) and Innovative Areas (20102005), and by Tohoku University, Inter-university Cooperative Research Program of the Institute for Materials Research. 

\bibliography{GeCo2O4_9_PRB}

\end{document}